%% file: author-start-here.tex
\documentclass[graybox]{svmult}
\usepackage{wrapfig}
\usepackage[numbers]{natbib}

\usepackage{amsmath}
\usepackage{algpseudocode}
\usepackage[ruled]{algorithm2e}
\usepackage{subfigure}
\usepackage{pifont}
\usepackage{array}
\newcolumntype{P}[1]{>{\centering\arraybackslash}p{#1}}

\usepackage{helvet}         
\usepackage{courier}        
\usepackage{type1cm}        
\usepackage{makeidx}         
\usepackage{graphicx}        
\usepackage{multicol} 
\usepackage[parfill]{parskip}
\usepackage[bottom]{footmisc}

\makeindex             

\begin{document}                     


\title*{GI-OHMS: Graphical Inference to Detect Overlapping Communities}
\author{Nasheen Nur, Wenwen Dou, Xi Niu, Siddharth Krishnan and Noseong Park}
\institute
{Nasheen Nur \at Unversity of North Carolina at Charlotte, North Carolina, \email{nnur@uncc.edu}
\and Wenwen Dou \at Unversity of North Carolina at Charlotte, North Carolina, \email{wdou1@uncc.edu}
\and Xi Niu \at Unversity of North Carolina at Charlotte, North Carolina, \email{xniu2@uncc.edu}
\and Siddharth Krishnan \at Unversity of North Carolina at Charlotte, North Carolina, \email{skrishnan@uncc.edu}
\and Noseong Park \at George Mason University, Fairfax, Virginia, \email{npark9@gmu.edu}}
%
%
\maketitle
\vspace{-8mm}
\abstract{Discovery of communities in complex networks is a topic of considerable recent interest within the complex systems community. Due to the dynamic and rapidly evolving nature of large-scale networks, like online social networks, the notion of stronger local and global interactions among the nodes in communities has become harder to capture. In this paper, we present a novel graphical inference method - GI-OHMS (Graphical Inference in Observed-Hidden variable Merged Seeded network) to solve the problem of overlapping community detection. The novelty of our approach is in transforming the complex and dense network of interest into an \emph{observed-hidden merged seeded}(OHMS) network, which preserves the important community properties of the network. We further utilize a graphical inference method (Bayesian Markov Random Field) to extract communities. The superiority of our approach lies in two main observations: 1) The extracted OHMS network excludes many weaker connections, thus leading to a higher accuracy of inference 2) The graphical inference step operates on a smaller network, thus having much lower execution time. We demonstrate that our method outperforms the accuracy of other baseline algorithms like OSLOM, DEMON,and LEMON. To further improve execution time, we have a multi-threaded implementation and demonstrate significant speed-up compared to state-of-the-art algorithms.}

\input{Introduction}


\input{Related}
\input{Methodology}

\input{Experiment}
\input{conclusion}
\bibliographystyle{unsrtnat}
\bibliography{bib}
\end{document}

%% file: Introduction.tex
\section{Introduction}
\label{sec:1}
\vspace{-4mm}
Communities in networks, particularly in large-scale social and information networks, help emphasize the relational nature of the complex system. The nodes within these communities display a dense set of interactions and remain weakly connected to nodes outside their communities. For example, a hashtag on Twitter, say \emph{\#fifaworldcup} will find a retweeting substructure among fans of football. A cluster of nodes in a metabolic network may represent a pathway or a cycle that is of interest to a biochemist. Using scientific citation networks, we find that scientists belonging to the same research community often collaborate. Extracting such communities from a network has diverse applications in sociology (organizational units in a social network), computer science (topically connected web pages), biology (functional units in biochemical networks)~\cite{ahuja2017communities}. Traditionally, in simple networks, unearthing community structure often partitions the graph into dense clusters. However, modern networks have a lot more complexity, and a single node can belong to more than one community. For example, an individual can belong to several groups in \emph{Facebook} and have different interactive patterns on those groups. Therefore, it is highly likely that a node in social network shares multiple interests and so belongs to multiple communities resulting in memberships in overlapping communities. In our work, we present a novel method aided by graphical inference to discover these overlapping collaborations in communities.

We formulate the problem of community detection as one of graph clustering problems. We present a graphical model approach, where we infer the community memberships of a node which is the random variable of interest. GI-OHMS outperforms the state-of-art algorithms where we consider discovery of local communities as a starting point and then expand our analysis by optimizing energy score and marginal probabilities in the global space. 

Following is a summary of our contribution-
\begin{itemize}
\item The main contribution of our paper is a new overlapping
community detection algorithm which greatly exceeds the performance of other state-of-the-art algorithms in terms of execution time, coherence of communities, and ground-truth accuracy.
\item We incorporate a new concept of seed expansion via graphical inference. The general approach for community detection is conducting a greed search on a modular structure of a network and recursively evaluating the modules via an objective function to return the best communities. Most of the time this approach fails due to complex and different modular structures locally and globally . To solve this difficulty of making a cohesion among global and local structure we propose a method which uses the local modular information and expand this local information by graphical inference in the whole network.
\item We implement a multi-threaded version to minimize the execution time overhead, where seed expansion to derivation of marginal probability of community variables are independent of each other.
\end{itemize}

The remainder of the paper is organized as follows. Section 2 gives the general overview of the ongoing research in this field. We introduce our methodology in  details in Section 3 followed by experimental setup and results analysis in Section 4.This paper concludes with future directions of research in Section 5.\vspace{-4mm}

%% file: Related.tex
\section{Related Work}
\label{sec:2}
\vspace{-4mm}
Over the last decade, for overlapping community detection, researchers applied many different approaches including clique
percolation, link partitioning, statistical inference, seed expansion, ego network analysis and fuzzy detection ~\cite{xie2013overlapping}.
Clique percolation~\cite{kumpula2008sequential,reid2012percolation,gregori2013parallel} looks for overlapping nodes among the
fixed size cliques in the graph whereas link partitioning~\cite{ahn2010link} (also known as line graph partitioning) introduces partitioning of the line graph to find out overlapping communities in the original graph. Clique percolation and link partitioning often fail to scale to large networks even though they are very effective methods to find overlapping modular structures in smaller graphs.

A soft clustering scheme applied to eigenvectors
of the normalized Laplacian or modularity
matrix in order to estimate communities in eigenvector methods~\cite{zhang2007identification}. Statistical ~\cite{macropol2011content,gopalan2013efficient} inference methods utilize both information content and graph structure to infer further information flow in overlapping community structure.Ego
network analysis methods use the theory of local communities and structural hole and compute overlapping communities by combining them~\cite{coscia2012demon}.We compare our algorithm with the Demon~\cite{coscia2012demon} method since we also introduce the local communities by ego network expansion.\\
The approach we employ is called local-first seed expansion.Seed expansion methods optimize the objective function by greedily searching in the network space. Defining a method to seed initialization is a crucial stage to this kind of approach. For example, OSLOM~\cite{lancichinetti2011finding} starts from a randomly
picked node, and then greedily expands the cluster by checking whether the expanded community is statistically significant or not.LEMON~\cite{li2015uncovering} expands the seeds with one norm where seeds are defined by fraction of true communities. We compare our method with both LEMON and OSLOM method in our experiments.

A qualitative comparison is provided in Table~\ref{comp} among different overlapping seed expansion methods and GI-OHMS.\vspace{-\parskip}
\begin{table*}
\caption{\textbf{Qualitative comparison of prior works and GI-OHMS}}
\centering

\label{comp} 

\begin{tabular}{|p{2in}|P{0.59in}|P{0.55in}|P{0.55in}|P{0.7in}|}\hline
\textbf{Properties} & \textbf{DEMON} & \textbf{LEMON} & \textbf{OSLOM} & \textbf{GI-OHMS}\\\hline
Parallelizable  &\ding{55}  &\ding{51}   &\ding{55}  & \ding{51}\\ \hline
Random initialization does not imply different results &\ding{55} &\ding{55}&\ding{55}  &\ding{51}\\ \hline
Does not require initialization with partial ground truth or state-of-art algorithms &\ding{51}&\ding{55}&\ding{55}  &\ding{51}\\ \hline
Scaleable for smaller to larger complex networks&\ding{55}  &\ding{51}   &\ding{55}  & \ding{51}\\\hline
Considers local \& global neighborhood sturcture for energy minimization &\ding{55}  &\ding{55}   &\ding{55}  & \ding{51}\\
 \hline

\end{tabular}
\end{table*}

%% file: Methodology.tex
\section{Methodology}
\vspace{-4mm}
Our goal in this paper is to find out overlapping communities in simple to complex social networks by utilizing and expanding neighborhoods' information for each vertex. \vspace{-5mm}
\subsection{Problem Formulation}%
\vspace{-5mm}
\textit{\textbf{Given}}: A social network represented as $G(V,E)$, with a vertex set $V$ and an edge set $E$.\\
\textit{\textbf{Goal}}: To find overlapping clusters whose union is not necessarily equal to the entire vertex set $V$. Formally, we seek $n$ overlapping clusters such that $C_1 \cup C_2 \cup..... C_n \subseteq V$. Communities capture the notion of stronger interactions among the nodes within the community than outside the community. In overlapping community detection, a node may belong to several communities. 
\vspace{-10mm}
\subsection{Preliminaries} %
\vspace{-5mm}
In this section, we discuss each component of our algorithm to solve the problem of overlapping community discovery. \vspace{-5mm}
\subsubsection{The Ego-Minus-Ego}
\vspace{-5mm}
The concept of ego network was first introduced by~\citet{burt1980models}. In an un-directed social network $G=(V,E)$, the \textit{ego network} of a vertex $v \in V$ is an induced sub-graph of a set $\{v\} \cup \{u | u\textrm{ is a neighbor of } v\}$, where $v$ is called \textit{focal vertex} (or \textit{ego vertex}) and $u$ is called \textit{alert}.
 \begin{definition}
 An induced sub-graph of a vertex set is a sub-graph $s$ of the original network $G$ such that 
\begin{itemize}
    \item $s$ has the exactly same vertices as the set and
    \item There is an edge $(x,y)$ on the sub-graph if and only if $(x,y)$ exists on the original network $G$
\end{itemize} \end{definition} \vspace{-3mm}
The \textit{ego-minus-ego network} of $v$ is a simple tweak from its ego network as follows.
\begin{definition}
Given an un-directed social network $G=(V,E)$, the \textit{ego-minus-ego network} of a vertex $v$ is defined as the induced sub-graph only by $\{u | u\textrm{ is a neighbor of } v\}$. Since the ego vertex (or focal vertex) is removed from the ego network, it is called ego-minus-ego network. 
\end{definition}

\textit{\textbf{Why we are removing ego node from its ego network?}}
\begin{itemize}
    \item We are developing a local first approach where each node will look for its own local community information. All edges in the ego network of a focal vertex $v$ are \emph{strong relationships} in the perspective of $v$ because those edges connect members of $v$'s cliques,which may not be case in the global communities. Therefore to avoid the bias towards only ego node's local community, we are building the ego-minus-ego network.
    \item \textbf{\textit{Label propagation}} is one of the fastest methods to find out local communities with a \textit{quasi-linear} time complexity. It's a local first method, where each node interacts with its neighbourhood to change its own label. Therefore,firstly,keeping the triangle with ego node while label propagation will rarely give same results for different starting point. Secondly, it may introduce noise since ego node will be connected to all other nodes in this sub-graph leading other nodes being in the same community even though they are not densely connected. Therefore, it will result in huge communities than actual situation.
\end{itemize}

In Figure 1(a), a toy social network is shown. Figure 1(b) is the ego network of D. Note that D is the ego vertex (or focal vertex). Figure 1(c) represents the ego-minus-ego network of D where D is removed from the network.\vspace{-4mm}

\begin{figure*}[t]
\centering

\subfigure[]{\includegraphics[width=0.25\textwidth]{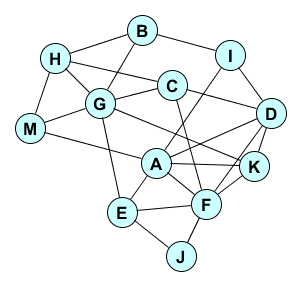}}
\subfigure[]{\includegraphics[width=0.25\textwidth]
{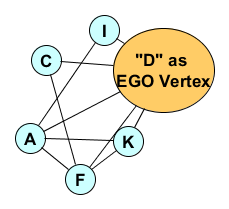}}
\subfigure[]{\includegraphics[width=0.15\textwidth]
{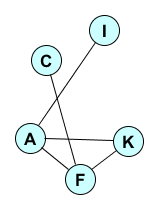}}
\subfigure[]{\includegraphics[width=0.25\textwidth]
{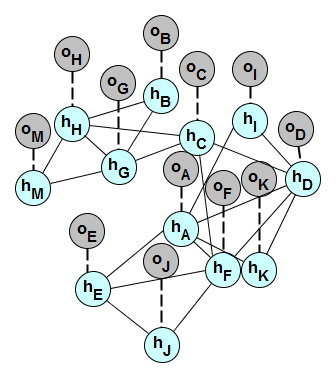}}
\caption{(a) Original network (b) The ego network of D (c) The ego-minus-ego network of D (d) OH-MSEN network to perform graphical inference. Note $h_X$ means a hidden variable that represents a real community membership of a vertex $X$ and $o_X$ is an observed variable.}\label{figure}\vspace{-4mm}
\end{figure*}
\subsubsection{Label Propagation Phase and Merged Seeded Network }
\vspace{-4mm}
We perform Label Propagation(LP)~\cite{raghavan2007near} in each EME network. This algorithm returns local communities for each of the node in the network. In every iteration ($t$) of the propagation, for each node  $x \in X$ where $X$ represents the set of randomly ordered nodes in the network and function $f$  returns the label occurring with the highest frequency among neighbours, such as  
\begin{equation}
    C_x(t) = f(C_{x_{i1}}(t - 1), ...,C_{x_{im}}(t - 1),C_{x_{i(m+1)}} (t − 1), ...,C_{x_{ik}}(t − 1))\vspace{-2mm}
\end{equation} 

The propagation process reaches its stopping criteria whenever it hits a maximum iteration or every node in the network has a label with a neighborhood of maximum nodes.\vspace{-2mm}
 
Then, the local communities are merged to the bigger set of communities by a threshold $\epsilon$. The threshold defines the percentage of absent communities in the smaller community among two communities. We call the merged network Meged Seeded Ego Network (MSEN). We assume community labels in this merged network are the initial seeds for seed expansion process.\vspace{-5mm} 

\subsubsection{Observed-Hidden Variable Network}
\vspace{-4mm}
MSEN is further modified and converted into an \textbf{\textit{Observed-Hidden Merged Seeded Ego Network} (OH-MSEN)}. We use the terms \textit{hidden variable} and \textit{observed variable} to distinguish two different types of variables. In an OH-MSEN, a vertex is called \textit{hidden variable} which has the same connection as the original network. Each \textit{hidden variable} has one more special neighbor variable representing an outside observation (See Figure 1 (d)). An \textit{observed variable} is the community label obtained by the label propagation. Therefore, an observed variable contains a certain value representing a community membership that is actually an observation from its neighboring hidden variables.\vspace{-4mm}
\subsubsection{Seed Expansion Via Graphical Inference}
\vspace{-4mm}
Hidden variables contain real community memberships and we consider seeds(observed variables) are actually observations for them.On OH-MSEN, we infer the values of the hidden variables by graphical inference to find out the final overlapping community memberships.\\
Almost all graphical inference methods commonly use the concept of \textit{energy}. An energy (or uncertainty) value can be defined for a complete set of hidden variable value assignments; and if a set of assignments are correct, then its energy value should be low. The most common energy definition is as follows.\vspace{-2mm}
\begin{footnotesize}
\begin{equation}
energy(O,H) = \sum_{i}UnaryCost(o_i, h_i) + \sum_{(i,j) \in E}{BinaryCost(h_i,h_j)}
\vspace{-2mm}
\end{equation}
\end{footnotesize}where $O$ is a set of observed values and $H$ is a set of hidden variable value assignments.
Both of $UnaryCost(\cdot)$ and $BinaryCost(\cdot)$ span in $[0,w]$, where $w$ is a user-defined max energy penalty. $w$ can be differently set to two different cost definitions.\vspace{-3mm}
\begin{definition}
The \textbf{\textit{unary cost}} checks if a hidden variable $h_i$ is the same as its observed variable $o_i$ (i.e., $UnaryCost(o_i, h_i)=0$ if $o_i = h_i$). In other words, we trust that outside observations are correct in many cases and this is the reason why we have to initialize observed variables with reliable local communities.\vspace{-3mm}
\end{definition}
\begin{definition}
The \textbf{\textit{binary cost}} requires that two neighboring hidden variables' community memberships are identical. \vspace{-3mm}
\end{definition}
 
We choose $h_i \neq o_i$ or $h_i \neq h_j$ only if they decrease the overall energy level. There are several energy minimization techniques. We use \textit{Bayesian Markov random field (BMRF)}~\cite{kourmpetis2010bayesian,Welling06bayesianrandom} for its very accurate inference performance. \\
BMRF lets $w$ vary for each $h_i$ in the \textit{unary cost} and each pair $(h_i,h_j)$ in the \textit{binary cost}. Thus, $w$ is actually a $d$-dimensional vector of parameters, where $d$ is the sum of the number of hidden variables and the number of edges between hidden variables. Its freedom of degree is maximized in the parameter setup.
We assume a prior on $w$. Many parameter samples are drawn from its posterior distribution via Gibbs or MCMC sampling including the posterior for $w$. We used an adaptive MCMC sampling called \textit{differential evolution Markov chain} (DEMC)~\cite{ter2008differential} to update the parameters in each step.
To have a robust inference, we perform an ensembling procedure of samples by taking an average inference value for each hidden variable.  Finally, communities are extracted with a higher joint probability (80\% to 100\%) for the hidden variables.\\
The following pseudo-code provides the overall summary for the algorithm.

\begin{algorithm}[H]
 \caption{Overlapping Community Detection}
     \KwData{Given a network $G:(V,E)$, a merging threshold $\epsilon \in [0\dots1]$, empty $Com$ set, community probability $p\in [0\dots1]$}
     \KwResult{a set of overlapping communities}
     \For{each vertex $v$}{
        Extract egoMinusEgo network $e_v$\;
        Extract seeded network, $C_v\Leftarrow labelPropagation(e_v$)\;
        \For{each community $C_i \in C_v$}{
            Add $v$ to $C_i$\;
            Merge $C_i$ to $Com$ by $\epsilon$\;
            }
        }
        Observed Hidden Merged Seeded Ego Network $OH-MSEN \Leftarrow Com$ \;
        
    \end{algorithm}
    \vspace{-4mm}

%% file: Experiment.tex
\section{Experiments \& Results}\vspace{-4mm}
\subsection{Data-sets}\vspace{-5mm}
We tested our algorithm on two real world complex social networks\footnote{http://snap.stanford.edu/data} provided with ground truth communities and five synthetic networks  of varying configuration and complexities developed by the famous Lancichinetti-Fortunato-Radicchi (LFR)~\cite{lancichinetti2008benchmark} community detection benchmark.
We selected the real-world network from different domain: Amazon from product domain and DBLP from collaboration domain. A general overview on the statistics of these networks can be found in Table~\ref{tab:data}.\vspace{-4mm}

\begin{table}
\centering
\caption{\textbf{Statistics of Data-sets}}
\label{tab:data} 
\resizebox{\textwidth}{!}{
\begin{tabular}{|c|c|c|c|c|c|}\hline
\textbf{Domain} & \textbf{Name} & \textbf{No. of Vertices} & \textbf{No. of edges} & \textbf{Max Degree} & \textbf{Average Degree} \\\hline
Product  & Amazon & 334,863 & 925,872 & 549 &5.5\\ \hline
Collaboration & DBLP & 317,080& 1,049,866& 343 & 6.6\\ \hline \hline \hline
Synthetic &Example 01&2000&13980  &50&15\\ \hline
Synthetic &Example 02&3000&23429  &60&20\\ \hline
Synthetic &Example 03&4000&38806  &80&25\\ \hline
Synthetic &Example 04&10000&138852 &80&45\\ \hline
Synthetic &Example 05&50000&930456 &100&50\\ \hline
\end{tabular}}
\vspace{-6mm}
\end{table}\vspace{-6mm}
\subsection{Evaluation Metrics}\vspace{-5mm}
We used ONMI and F1-score to evaluate the performance of our community detection algorithm. Given a set of true communities, and the set of communities found by an algorithm, later communities must be compared to see how similar or different the sets are.\vspace{-8mm}
\subsubsection{Overlapping Normalized Mutual Information(ONMI)}\vspace{-5mm}
Given two set of communities X and Y have mutual information I, and H(X) and H(Y) are the marginal entropy respectively, therefore ONMI will be defined as~\cite{mcdaid2011normalized}
\begin{equation} \label{eq1}
\begin{split}
ONMI = \frac{I(X:Y)}{max(H(X),H(Y)))} 
\end{split}
\end{equation}
ONMI is a measure of the inherent dependence which quantifies the joint distribution of X and Y according to the joint distribution of X and Y under the assumption of independence. If X and Y have zero mutual information, the above equation implies that OMNI will be also zero.\vspace{-8mm}

\subsection{Experimental Results}
\vspace{-5mm}
We compared GI-OHMS with three other state-of-art seed expansion methods : DEMON~\cite{coscia2012demon}, OSLOM~\cite{lancichinetti2011finding} and LEMON~\cite{li2015uncovering}. Table~\ref{f1} summarizes the execution time as well as the average F1 score of each algorithm on real datasets. Among the state-of-art algorithms, even though OSLOM achieved a comparatively good results than other algorithms, it failed to scale well for larger data-sets. It takes days(with 8GB RAM processor) to compute communities in the larger real-world datasets. Although, BMRF introduces complexity in computation, our multi-threaded implementation achieved the communities with the comparatively best result in a shorter time span. We further analyze the results in the conclusive part of this section.
\vspace{-5mm}
\begin{table}
\centering
\caption{\textbf{Comparison of Algorithms in real data-sets}}
\label{f1} 
\resizebox{\textwidth}{!}{
\begin{tabular}{|c|c|c|c|c|c|c|}\hline
 & \multicolumn{3}{c|}{\textbf{Amazon}}& \multicolumn{3}{c|}{\textbf{DBLP}}\\\hline\hline
\textbf{Algorithm} & \textbf{F1-Score} &\textbf{Execution time} & \textbf{\# of Threads} & \textbf{F1-Score} &\textbf{Execution time} & \textbf{\# of Threads}\\\hline
OSLOM & 0.720 & $\approx 6 days$ & Single & 0.492 &   $\approx 11 days$ & Single\\\hline
DEMON  & 0.156 & $\approx 6 hours$ & Single & 0.177 &  $\approx 4 days$ & Single\\ \hline
LEMON & 0.932 & $\approx 20 seconds$ & Single & 0.710 &  $\approx 20 seconds$ & Single\\ \hline \hline \hline
GI-OHMS & 0.967 & $\approx 2 hours$ & 12 & 0.829 &  $\approx 18 hours$ & 16 \\ \hline

\end{tabular}}

\end{table}
The multi-threaded version of  our  algorithm  has  small  memory  consumption,minimized system resource usage, and since the local seed expansion is parallelizable and independent of each other, it supports simultaneous and fully symmetric use of multiple processors for faster computation. This property
brings performance gain on execution time with multi-threaded implementation.

\begin{figure*}[t]
\centering

\subfigure[]{\includegraphics[width=0.45\textwidth]{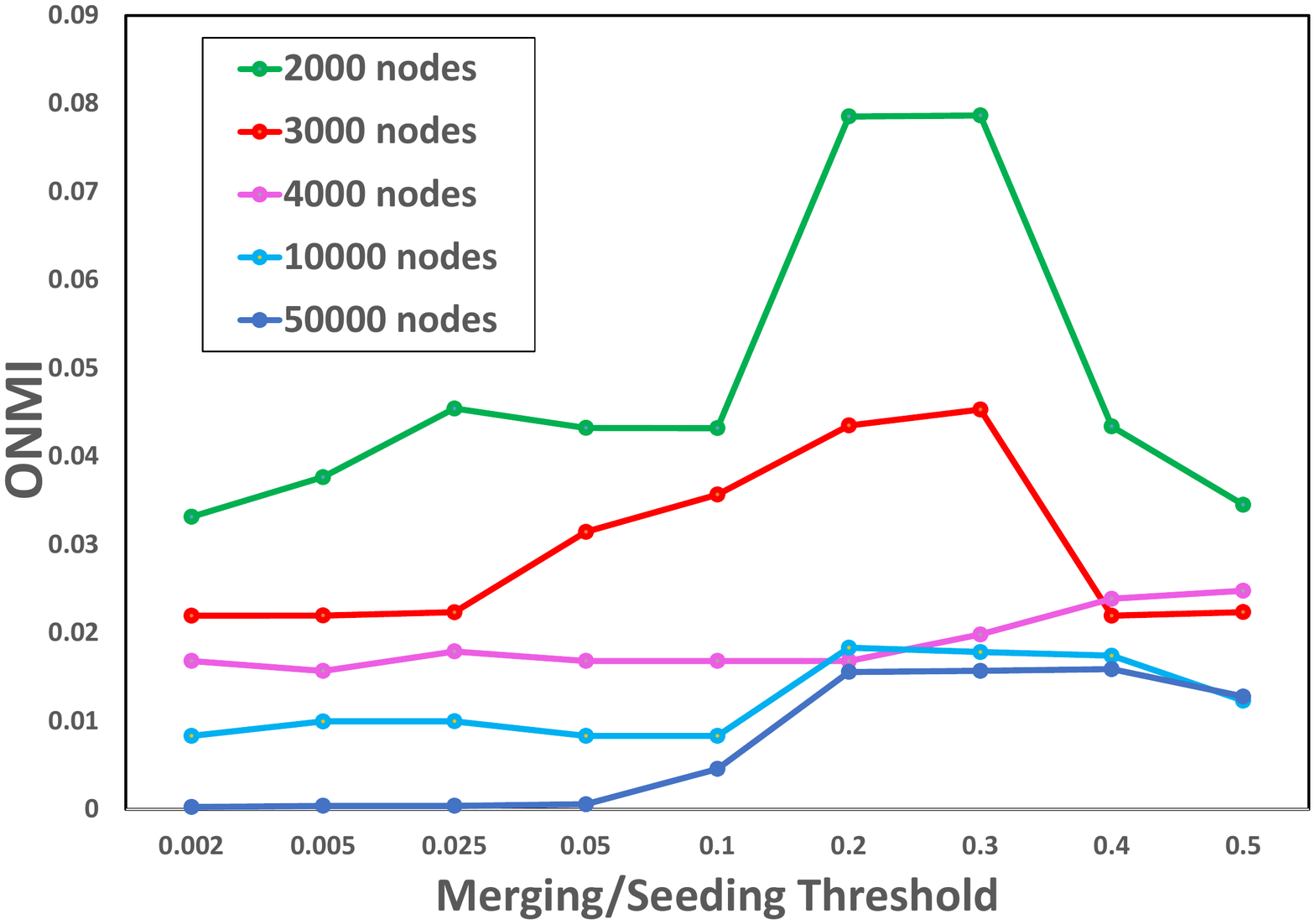}}\label{onmi}
\subfigure[]{\includegraphics[width=0.45\textwidth]{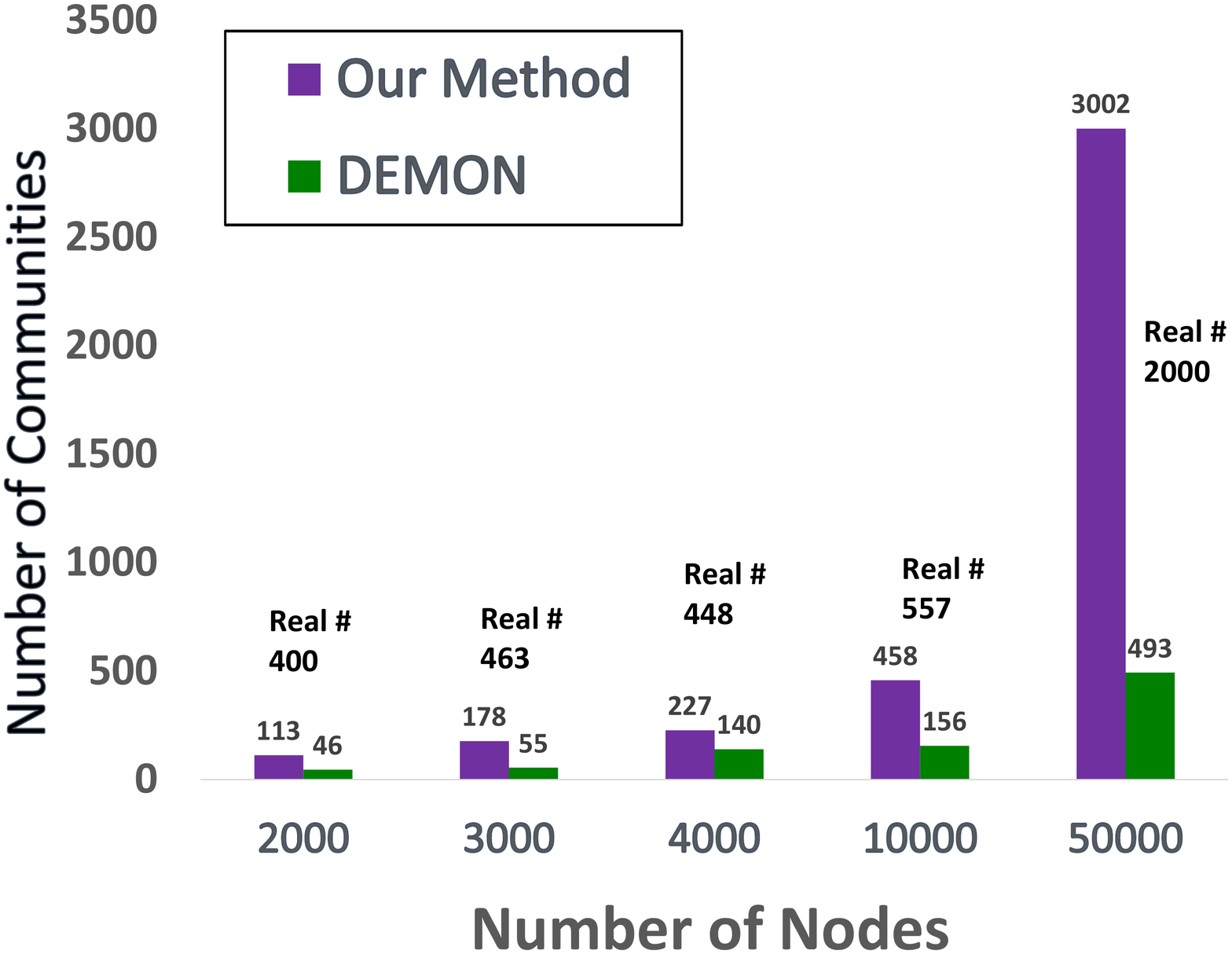}}\label{onmi2}

\caption{(a) ONMI with varying threshold for 5 benchmark data-sets (b) Communities numbers in DEMON Vs GI-OHMS (for $\epsilon = 0.1$)}\label{figure2}
\vspace{2mm}
\end{figure*}
Figure 2(a) compares the ONMI for different size of LFR benchmark graphs with varying threshold. We observe that the performance is independent of the size of the network. Later in the Table~\ref{onmitbl} we show that we achieve a far better ONMI than the state-of-art algorithms on the benchmark data-sets. 
\begin{table}
\centering
\caption{\textbf{Comparison of ONMI in Benchmark data-sets}}
\label{onmitbl} 
\resizebox{\textwidth}{!}{
\begin{tabular}{|c|c|c|c|c|c|}\hline
 \textbf{Algorithms} & \textbf{With 2000 node} &\textbf{With 3000 node} &\textbf{With 4000 node} &\textbf{With 10000 node} & \textbf{With 50000 node}\\\hline

OSLOM &0.0006321&0.007341&0.008953&0.0032168&0.00095412\\\hline
DEMON  &0.00940351&0.0148336&0.0180243&0.00478648&0.003132\\ \hline
LEMON &0.021543&0.0112&0.0159&0.00634251&0.0053212\\ \hline \hline \hline
GI-OHMS &0.0331584&0.0210037&0.0167241&0.00897371&0.00655632 \\ \hline

\end{tabular}}
\vspace{-4mm}
\end{table}
We expanded our algorithm from the DEMON implementation with graphical inference by energy minimization. After the merge operation in the local communities DEMON is directly outputing the communities without any further global operation resulting to increasing the chance to local minima. Therefore, DEMON usually finds communities that are much larger than the ground truth communities which implies that communities which has lesser resemblance to the original topology of the communities. It is noticeable that our algorithm outperforms the DEMON algorithm even when we use varying threshold to identify the actual number of communities (See Figure 2(b)). However, selecting a appropriate seeding threshold is critical to the presented algorithm.

Following is a summary of the features of the presented algorithm.

\begin{itemize}
\item Firstly, we observe in Table~\ref{f1} that \textit{LEMON} has far better execution time than \textit{GI-OHMS}, however, \textit{LEMON} performs better while initialized by a portion of ground truth community information which is biasing the output of this algorithm. For the comparative analysis on \textit{LEMON} and \textit{GI-OHMS}, we initialized \textit{LEMON} with max possible grount-truth community information. On the other hand, \textit{OSLOM} uses other state-of-art algorithms (i.e Infomap) in their initialization process. Our  \textit{GI-OHMS} initializes the seeds by looking into the node similarity via label propagation. Therefore, it is independent and neither biased by other community detection algorithms nor the original ground truth labels. 
    \item Secondly, \textit{GI-OHMS} is considering both local and global topology of a network. It is removing the bias of local communities by assigning marginal probabilities over the whole neighborhood (closer to distant) by energy minimization.
    \item Thirdly, it is less sensitive to the size of the network. \textit{DEMON} and \textit{OSLOM} performs worse when input network is larger and has a higher average degree. 
    \item \textit{GI-OHMS} provides a scalable framework within the scope of independent and parallel seed expansion.
   
    \item Finally, the percentage of seeded information directly influences the method although the effect is later minimized by global graphical inference (See Figure 2(a)). This can be a improvement scope for the future research.
\vspace{-5mm}    
\end{itemize}


%% file: conclusion.tex
\section{Discussion \& Conclusion}
\vspace{-3mm}
In this paper, we have presented a method for finding overlapping communities by expanding sparser local communities from a single node point view to global space of a network. We present a novel method to identify the global communities via seed expansion in local-node level communities with the energy minimization concept of graphical inference. It also enables the scope for parallel computing to handle bigger and higher degree networks.  We use both the synthetic and real-world datasets to justify the performance efficiency comparing to the state-of-art algorithms. We are scoping out several other research questions to leverage future research direction on the presented framework. Those scopes could be newer strategies to find out local communities, applications of ego-minus-ego network in other graph analysis,and analysis of behavioral trend from local to global communities.\vspace{-6mm}